\documentclass[a4paper,11pt]{article}
\usepackage{pos}

\title{Hunting for Dark Photon-Photon Tridents}

\author*[a]{Thong T.Q. Nguyen}

\affiliation[a]{Stockholm University and The Oskar Klein Centre for Cosmoparticle Physics\\
  Alba Nova, 10691 Stockholm, Sweden}

\emailAdd{thong.nguyen@fysik.su.se}

\abstract{
Dark photons, hypothesized to be sufficiently light and/or weakly interacting, offer a compelling candidate for dark matter. Their decay into three photons, referred to as the "dark photon trident" process, becomes the dominant channel when the dark photon mass lies below the electron pair production threshold. This decay channel produces a significant flux of x-rays, presenting an opportunity for indirect detection. In this study, we analyze 16 years of x-ray data from INTEGRAL/SPI to investigate sub-MeV dark photon decay. By incorporating state-of-the-art astrophysical background modeling and accounting for the full one-loop decay amplitude, we achieve world-leading constraints on the kinetic mixing parameter for dark photon masses in the range of 61--1022 keV. These results represent a significant improvement over previous constraints, narrowing the parameter space for viable dark photon dark matter models. Furthermore, our findings highlight the potential of x-ray observatories to probe unexplored regions of parameter space and pave the way for future searches using next-generation instruments designed to detect faint astrophysical signals.
}

\FullConference{2nd Training School and General Meeting of the COST Action COSMIC WISPers  (CA21106) (COSMICWISPers2024)\\
 10-14 June 2024 and 3-6 September 2024\\
Ljubljana (Slovenia) and Istanbul (Turkey)\\}

\begin{document}
\maketitle

\section{Introduction}
\label{sect:intro}
The dark photon, a hypothetical spin-1 particle from extensions of the Standard Model, interacts with fermions via kinetic mixing with the photon and is categorized as a Weakly Interacting Slim Particle (WISP)~\cite{Fabbrichesi:2020wbt}. It is a viable dark matter candidate if stable on cosmological timescales, though non-zero kinetic mixing allows decay into detectable Standard Model particles, producing bright photon signals observable in the electromagnetic spectrum.

For dark photon masses below twice the electron mass, only photon and neutrino final states are accessible~\cite{Nguyen:2022zwb}. Due to the Landau-Yang theorem forbidding two-photon final states and the suppression of decays to neutrinos, the dominant decay channel is the three-photon final state, known as the "dark photon trident," which produces a broader spectrum~\cite{Linden:2024uph}. While constraints rule out large kinetic couplings in the sub-MeV range, sensitivity gaps remain in the x-ray band~\cite{Caputo:2021eaa}. Detecting these diffuse signals is challenging, but wide-field-of-view telescopes like INTEGRAL are more effective than high-resolution instruments like Chandra or XMM-Newton.

In this work, we analyze 16 years of x-ray data from INTEGRAL/SPI to impose stringent constraints on dark matter decays via photon tridents in the mass regime where the dark photon is below twice the electron mass~\cite{Linden:2024fby}. By incorporating the one-loop electron enhancement in the dark photon decay rate, our results surpass previous analyses that relied on the leading Euler-Heisenberg approximation~\cite{Redondo:2008ec}. These findings strongly motivate further dark photon searches using next-generation x-ray telescopes.

\section{X-ray signals from dark photon trident and background model}
\label{sect:DP}

Assuming the dark photon constitutes the entirety of dark matter in the universe, we calculate the photon flux produced by the unique trident process $A^{\prime}\to 3\gamma$ in the Galactic Center region. The decay width of this channel is determined while incorporating the one-loop electron correction, which significantly enhances the decay width near the threshold where the dark photon mass approaches twice the electron mass. This enhancement can increase the decay width by up to two orders of magnitude, as shown in previous studies~\cite{McDermott:2017qcg}.

For the INTEGRAL/SPI x-ray signal from this dark matter decay, we follow the framework from Ref.~\cite{Calore:2022pks} to calculate the total flux. We consider the region of interest with $|l|\leq 47.5^{\circ}$ and $|b|\leq 47.5^{\circ}$ in the Galactic halo. Using the NFW profile with $r_{s}=9.98$~kpc and $\rho_{\odot}=0.42$~GeV/cm$^{3}$, we obtain $D=1.3\times10^{23}$~GeV/cm$^{2}$ for this region. We also remove the isotropic portion of the dark matter decay signal, leaving us with an effective $D$-factor of $D=1.2\times 10^{23}$~GeV/cm$^{2}$.

For the INTEGRAL background signal, we consider the background model that includes four major components~\cite{Calore:2022pks, Linden:2024fby}: (1) unresolved point sources modeled with an exponential-cutoff power-law, (2) inverse Compton scattering of cosmic ray electrons with the interstellar radiation field, characterized by a power-law spectrum, (3) positronium decay contributing to the 511~keV line, and (4) the $^{7}$Be nuclear line at 478~keV. Using INTEGRAL data for photon energies in the range 30--511~keV and the \texttt{3ML} package~\cite{Foreman-Mackey:2012any}, we implement all these signal and background components to constrain the dark photon trident decay process.

\section{INTEGRAL/SPI constraint on dark photon dark matter}
\label{sect:constrant}

\begin{figure}[t]
\centering
\includegraphics[width=0.6\textwidth]{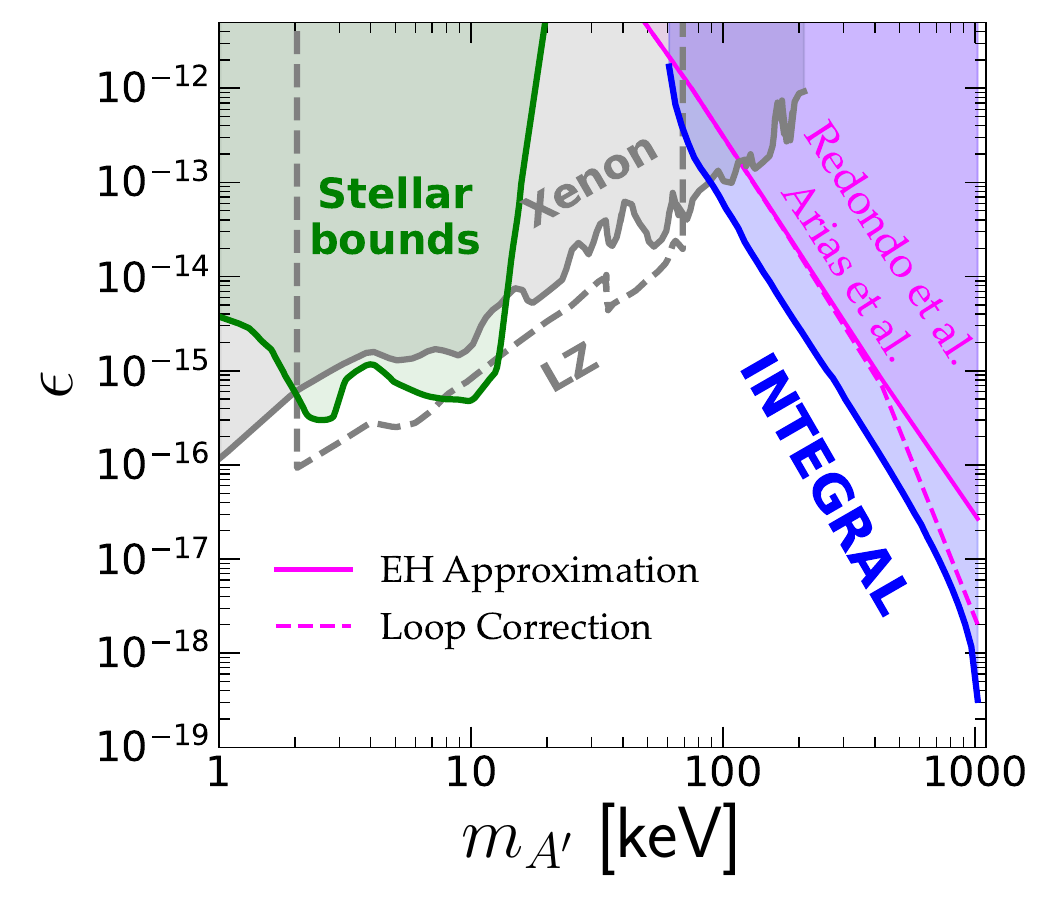}
\caption{Constraints on the kinetic mixing and dark photon mass from INTEGRAL/SPI are shown in blue. Previous limits from the diffuse x-ray background appear in magenta, with solid and dashed lines representing the EFT and one-loop results, respectively. Gray and green lines show direct search and stellar bounds, respectively.}
\label{figure}
\end{figure}

Figure~\ref{figure} shows our constraints for the dark photon dark matter model using INTEGRAL, with 95\%~CL (blue)~\cite{Linden:2024fby}. These constraints, shown as the blue line, significantly improve upon previous studies using the diffuse x-ray background, represented by magenta lines. Combined stellar production constraints, shown in green, provide complementary limits, particularly for low-mass dark photons, without assuming that dark photons constitute all dark matter. Our analysis demonstrates the unique power of diffuse x-ray data to probe dark matter mass ranges inaccessible to other methods and highlights the role of improved background modeling, the inclusion of the one-loop decay amplitude, and extended data sets, which together enhance sensitivity by an order of magnitude.

Our study focuses on dark photon masses below the electron-positron pair threshold, where the dominant decay channel produces photons without generating electrons or positrons. This allows us to leverage the INTEGRAL energy range to set robust constraints. By addressing astrophysical backgrounds and refining decay width calculations, we provide stronger and more reliable constraints compared to earlier works. These advancements underline the importance of precise modeling of astrophysical backgrounds in dark matter searches and motivate further studies to explore higher-mass regimes using the same approach.

\section{Conclusions}

In this study, we analyzed 16 years of INTEGRAL/SPI data to constrain the decay of dark photon dark matter in the 61–1022 keV mass range, focusing on the three-photon final state (dark photon trident) as the dominant decay channel. By incorporating astrophysical background modeling and the full one-loop decay amplitude, we achieved significant improvements over previous constraints, narrowing the kinetic mixing coupling limits by nearly two orders of magnitude. These findings not only strengthen our understanding of dark photon properties but also highlight the potential for extending this approach to broader searches for weakly interacting slim particles (WISPs) using existing and future x-ray and infrared observations~\cite{Nguyen:2024kwy}.

\section*{Acknowledgements}
TTQN is supported in part by the Swedish Research Council under contract 2022-04283, and by the Swedish National Space Agency under contract 117/19. This article is based on the work from COST Action COSMIC WISPers CA21106, supported by COST (European Cooperation in Science and Technology).

\end{document}